\documentclass[sigconf,nonacm]{acmart}

\usepackage{graphicx}
\usepackage{booktabs}
\usepackage{enumitem}
\usepackage{tikz}
\usepackage{xurl}
\usetikzlibrary{positioning, backgrounds, fit, calc}

\AtBeginDocument{
  }

\renewcommand{\showeprint}[2][]{\unskip}

\begin{document}

\title{To Ban or not to Ban? How Open Source Projects Govern GenAI Contributions}

\author{Wenhao Yang}
\orcid{0000-0002-1005-1974}
\affiliation{
  \institution{Peking University}
  \city{Beijing}
  \country{China}
}
\email{yangwh@stu.pku.edu.cn}

\author{Runzhi He}
\affiliation{
  \institution{Peking University}
  \city{Beijing}
  \country{China}
}
\email{rzhe@pku.edu.cn}

\author{Minghui Zhou}
\orcid{0000-0001-6324-3964}
\authornote{Corresponding author.}
\affiliation{
  \institution{Peking University}
  \city{Beijing}
  \country{China}
}
\email{zhmh@pku.edu.cn}

\begin{abstract}

Generative AI (GenAI) is playing an increasingly important role in open source software (OSS).
Beyond completing code and documentation,
GenAI
is increasingly involved in
issues, pull requests, code reviews, and security reports.
Yet, cheaper generation does not mean cheaper review — and the resulting maintenance burden has pushed OSS projects to experiment with GenAI-specific rules in contribution guidelines, security policies, and repository instructions, even including a total ban on AI-assisted contributions.
However, governing GenAI in OSS is far more than a ban-or-not question.
The responses remain scattered, with neither a shared governance framework in practice nor a systematic understanding in research.
Therefore, in this paper, we conduct a multi-stage analysis on various qualitative materials related to GenAI governance retrieved from 67 highly visible OSS projects. Our analysis identifies recurring concerns across contribution workflows, derives three governance orientations, and maps out 12 governance strategies and their policy instruments. We show that governing GenAI in OSS extends well beyond banning — it requires coordinated responses across accountability, verification, review capacity, code provenance, and platform infrastructure. Overall, our work distills dispersed community practices into a structured overview, providing a conceptual baseline for researchers and a practical reference for maintainers and platform designers.

\end{abstract}



\maketitle

\section{Introduction}

Generative AI is actively reshaping the future of OSS communities.
It is increasingly acting as a community contributor---submitting issues, pull requests, code reviews, and security reports~\cite{zhang2026survey,hou2024large,fan2023large}.
Recent coding agents take this further by navigating repositories, generating full patches, and executing pipelines with greater autonomy~\cite{yang2024swe,wang2024openhands,li2025aidev}.
With the rise of GenAI, how developers enter, participate in, and evolve community collaboration is undergoing a fundamental shift.

However,
 the unprecedented speed of generating contributions comes at a cost.
A prosperous and sustainable open source community depends not only on active contribution, but also on the capacity to review, triage, and maintain incoming work~\cite{gousios2014exploratory,zhou2019what,alami2022pull,wessel2023github,steinmacher2015systematic,eghbal2020working,linaker2024sustaining}.
As a maintainer of matplotlib observed, AI ``changes the cost balance between generating and reviewing code''~\cite{matplotlib_pr_31132}.
The surge of AI generated contributions grew beyond the maintenance capabilities of many well-known open source projects, causing chaos and debates within the cornerstones of the modern digital infrastructure.
The asymmetry can be stark: when a Node.js TSC member submitted a 19,000-line PR generated with Claude Code~\cite{nodejs_pr_61478}, the incident triggered a petition by over 80 developers~\cite{nodejs_ai_petition, nodejs_ai_petition_github} calling for a project-wide ban on AI-assisted contributions.

In response, projects have begun to establish GenAI-specific rules across contribution guidelines, security policies, issue and PR templates, AI directives (such as \texttt{AGENTS.md}), and public announcements.
The range of these responses is wide: some projects categorically reject AI-generated contributions on provenance and licensing grounds;
others require disclosure of AI use and impose accountability conditions on contributors;
still others constrain AI tool behavior directly through repository-level instructions;
and some have gone further, suspending external PRs or even migrating their primary collaboration venue.
At the same time, many projects acknowledge the productivity benefits and recognize that ``continuously rejecting LLM-generated code is not a good approach''~\cite{immich_discussion_23745}.
These diverse responses suggest that governing\footnote{Following prior work on OSS contribution governance~\cite{alami2022pull,elazhary2019guidelines,eghbal2020working}, we use \emph{governance} here to refer to the rules, procedures, and collaboration interfaces through which projects regulate contribution intake, review, accountability, and maintenance.} GenAI in OSS extends well beyond a binary question of permitting or prohibiting AI---yet current practices remain scattered across heterogeneous governance channels, with no shared framework or established best practice in sight.

This gap is both practical and academic.
Prior research has extensively studied GenAI's technical capabilities across software engineering tasks~\cite{zhang2026survey,hou2024large,fan2023large}, its adoption and effects on developer productivity and OSS collaboration~\cite{peng2023impact,baird2024early,metr2025measuring,fawzy2026vibe,zhao2026vibe}, and the long-standing governance mechanisms of OSS contribution management~\cite{gousios2014exploratory,zhou2019what,steinmacher2015systematic,eghbal2020working}.
However, these streams have not yet converged on the question of how OSS projects govern GenAI within contribution workflows.
We still lack a systematic understanding of what concerns GenAI raises across different stages of OSS contribution workflows, and how projects organize their responses into governance approaches and concrete strategies.

In this paper, we present the first systematic analysis of GenAI governance in OSS.
We collected repository-embedded governance materials (e.g., CONTRIBUTING, SECURITY, issue and PR templates, \texttt{AGENTS.md}) and surrounding justificatory materials (e.g., announcements, discussion threads) from 67 highly visible OSS projects.
Through iterative qualitative coding, we identified recurring concerns, governance orientations, and cross-project strategies.
Specifically, we answer the following research questions:

\textbf{RQ1.} \textit{What concerns does GenAI raise across OSS contribution workflows?}
\textbf{Result:} We identified seven recurring concern areas distributed across code contribution, communication, issue entry, security reporting, incentive structures, provenance and licensing, and platform infrastructure.

\textbf{RQ2.} \textit{What governance approaches do OSS projects develop for GenAI contributions?}
\textbf{Result:} We derived three governance orientations: prohibitionist, boundary-and-accountability, and quality-first---operationalized through 12 recurring governance strategies.

Our main contributions in this paper are:

\begin{itemize}[topsep=.1em, partopsep=.1em]
\item We provide a structured account of how GenAI-related concerns are distributed across OSS contribution workflows;
\item We derive three governance orientations that explain why projects facing similar GenAI pressures develop markedly different institutional responses;
\item We abstract a reusable strategy space of 12 governance strategies and their policy instruments, showing how the strategies serve different roles under different orientations.
\end{itemize}

Overall, our work offers both a conceptual baseline for future research on GenAI and software collaboration, and a practical reference for maintainers, communities, and platform designers seeking to establish or evaluate their own governance approaches.

\vspace{-.5em}

\section{Related Work}

In this section, we review prior work on: 1) GenAI for software engineering; 2) the adoption and influence of GenAI in the scope of OSS; 3) how OSS communities review, moderate, and maintain contributions.

\vspace{-.7em}

\subsection{GenAI for Software Development}

Since the rise of large language models,
we have observed and measured the effect of this technology in various software engineering tasks,
including code generation, testing, repair, and broader development assistance~\cite{zhang2026survey,hou2024large,fan2023large}.
Researchers have evaluated the code quality and found limitations of code completion products like GitHub Copilot~\cite{nguyen2022empirical}.
More recently, researchers have shifted focus to coding agents (e.g., SWE-Agent~\cite{yang2024swe}, OpenHands~\cite{wang2024openhands}) that navigate repositories, generate full patches, and execute pipelines with greater autonomy. Recent OSS-facing work also examines context-engineering practices and \texttt{AGENTS.md} files used to provide repository-specific instructions to such agents~\cite{mohsenimofidi2026context,gloaguen2026evaluating}. Through large-scale mining of agent-assisted pull requests, researchers~\cite{yang2024swe,wang2024openhands,li2025aidev}
investigate the status and
draw the technical boundaries of GenAI in software development.

\subsection{GenAI Use and Impact in Software and OSS Collaboration}

Researchers have also noted the influences of GenAI in other dimensions of software engineering, specifically in developer productivity, software quality, and collaboration.
Prior studies observed significant productivity gain from the use of GitHub Copilot~\cite{peng2023impact}, along with
higher validation and coordination costs in realistic development settings~\cite{baird2024early,metr2025measuring,forsgren2021space}. Surveys of AI coding assistants in practice report broad adoption, but also persistent trust and workflow concerns~\cite{sergeyuk2025using}. In OSS contexts, Xiao et al.~\cite{xiao2025self} examine self-admitted GenAI usage patterns across OSS repositories and, as a secondary contribution, qualitatively analyze 13 policy documents from 12 projects, classifying each by its observable stance toward GenAI as prohibitive, restrictive, or supportive; related studies analyze how developers self-declare AI-generated code and what such self-declarations contain~\cite{tufano2026developers,kashif2025self}. Other recent work has further explored vibe coding, Cursor AI, auto-merged agentic PRs, and human-AI review practices, revealing implications for vulnerability exposure, complexity, code survival, and review dynamics~\cite{fawzy2026vibe,zhao2026vibe,he2026speed,yoshioka2026agentic,branco2026lgtm,gao2026autopilot}. Related work also begins to surface compliance and provenance risks around AI-generated code, especially license uncertainty and compatibility in OSS settings~\cite{xu2024licoeval,wu2024large}, as well as tool support for provenance and license analysis of generated code~\cite{bifolco2025codegenlink}. Overall, this stream explains how GenAI is used and what effects it has on software work and OSS collaboration.

\subsection{OSS Contribution Governance, Maintainer Workload, and Moderation}

Our study builds most directly on prior work on OSS contribution governance and maintainer workload. Pull-based development, PR review, triage, and newcomer barriers have long been recognized as central issues in OSS collaboration~\cite{gousios2014exploratory,zhou2019what,steinmacher2015systematic}. More fine-grained studies have examined the governance interfaces through which projects manage these pressures: contribution guidelines often diverge from actual contribution processes~\cite{elazhary2019guidelines}, issue templates shape the processability of incoming reports~\cite{sulun2024issuetemplates}, and testing-related guidance in contribution documents remains uneven across projects~\cite{falcucci2025testingguidelines}. Related work has also shown that issue triage, maintainer scalability, duplicate submissions, review styles, automation infrastructure, and security reporting mechanisms are not merely coordination details, but important governance mechanisms for filtering inputs, allocating attention, and sustaining order in OSS projects~\cite{xie2013impact,zhou2017scalability,li2022duplicate,alami2022pull,wessel2023github,eghbal2020working,linaker2024sustaining,kancharoendee2025security}. This literature provides an important baseline for our work: the significance of GenAI lies not only in accelerating content generation, but also in intensifying and reshaping long-standing OSS problems around input filtering, accountability, review capacity, and maintainer overload.

\subsection{The Knowledge Gaps}
Existing work has mapped GenAI’s technical capabilities, developer adoption patterns, and individual productivity gains, while also establishing the importance of contribution governance and maintainer burden in traditional OSS.
The closest prior work is Xiao et al.~\cite{xiao2025self}, who study self-admitted GenAI usage in OSS and, as part of that broader analysis, classify 13 policy documents from 12 projects as prohibitive, restrictive, or supportive toward GenAI use. This provides an important first view of observable policy stance, but it leaves open a broader governance question: how do OSS projects define the GenAI problem, where do they encode governance rules, and through what mechanisms do they regulate GenAI-mediated contribution workflows? Our study addresses this question by analyzing 67 OSS projects across contribution guidelines, security policies, issue and PR templates, \texttt{AGENTS.md} files, and surrounding discussions. Rather than coding only a document's stance toward GenAI, we identify the concerns that motivate governance, derive the orientations that organize project responses, and abstract the concrete strategies through which those responses are operationalized. Going beyond previous findings on projects' positions toward AI contribution, our orientations encode the rationale from perception to action.
The distinction from previous stances is especially sharp in the
quality-first orientation (O3). O3 projects
reassert existing quality, reviewability, and maintainer-capacity thresholds instead of designing AI-specific rules --- not ``supportive''.

\section{Methodology}

\subsection{Research Design and Scope}

We conducted a comparative qualitative document analysis of publicly visible governance materials that OSS projects use to regulate GenAI-mediated contribution workflows. We did not aim to estimate policy prevalence or test a predefined theory. Instead, we sought to explain how OSS projects articulate GenAI-related problems, organize them into broader governance orientations, and express them through concrete governance strategies.

We therefore focused on \emph{publicly encoded governance}: texts that explicitly connect GenAI-related inputs to rules about admissibility, disclosure, accountability, verification, review, reporting, sanctions, or infrastructural arrangements. We treated a text as evidence of GenAI governance only when it explicitly regulated GenAI-assisted contribution behavior rather than merely mentioning AI, automation, or code quality. We fixed this inclusion criterion before document analysis began, building on prior definitions of OSS contribution governance~\cite{alami2022pull,elazhary2019guidelines,eghbal2020working}. The criterion stayed stable throughout screening; we only refined how it applied to borderline cases, so it was not shaped by our findings. Inclusion did not depend on source type: whether a rule appears in a wiki page, a blog post, or a markdown file explicitly marked for AI governance did not affect the decision. A source qualified when it was project-linked, maintainer-authored, or directly referenced by sampled governance materials and stated explicit rules for handling GenAI-assisted contributions; a CONTRIBUTING file or issue thread that merely mentioned AI did not.

Document analysis fits this goal because contribution guides, security policies, templates, AI policies, and tool-facing instructions are visible collaboration interfaces rather than post hoc commentary. The corpus comprises 67 OSS projects: 58 repositories and 9 communities. The communities are project-adjacent policy sources at the foundation, organization, or ecosystem level that directly govern OSS contribution practices. For consistency, we use the term ``OSS project'' for both kinds of cases throughout the paper.

\subsection{Corpus Construction}

To combine broad initial coverage with theoretically motivated follow-up sampling, we constructed the corpus in two stages.

\paragraph{Phase 1: broad repository screening.}
We created the seed frame from the top 800 starred repositories listed on Gitstar Ranking's repository leaderboard~\cite{gitstar_ranking_repositories}. This choice gave us a high-visibility set of projects with relatively mature governance surfaces and active external contribution workflows. We first removed non-code repositories (e.g., awesome lists and documentation-only collections) from the candidate pool, so no non-engineered repository entered screening. Two authors then independently examined each remaining repository's governance surfaces (CONTRIBUTING, SECURITY, standalone AI policies, \texttt{AGENTS.md}, and issue/PR templates) for public text that tied GenAI-related inputs to contribution rules, review expectations, security handling, channel restrictions, or sanctions. General references to AI, automation, or code quality were excluded unless they were explicitly connected to OSS contribution governance. Disagreements were resolved through discussion.
This stage yielded 37 OSS projects with explicit GenAI governance content.

\paragraph{Phase 2: snowball expansion.}
We then expanded the corpus through structured snowballing, following the iterative procedure formalized by Wohlin~\cite{wohlin2014guidelines}. Starting from the 37 seed projects, we retrieved and read the commits, issues, and PRs that added or modified the governance rules. For each relevant external reference---third-party policy pages, foundation governance texts, or AI policies of other projects---we manually validated it against the inclusion criterion and decided whether to expand the corpus. Along these paths we also captured governance surfaces underrepresented in the seed set, such as standalone AI policies, AI-specific security reporting rules, \texttt{AGENTS.md} files, and platform-migration announcements. Materials that merely mentioned AI without setting contribution rules were excluded at every stage. Expansion stopped when two criteria were met: \emph{connectivity saturation}, when new links no longer produced qualifying cases, and \emph{category saturation}, when new cases no longer introduced new codes. We deliberately opted out of random sampling because explicit GenAI governance rules remain rare in the open source world, and snowballing from a well-defined seed set is an established procedure for finding rare, interconnected artifacts. This stage added 30 cases and brought the final corpus to 67 OSS projects: 58 repositories and 9 communities. As shown in Table~\ref{tab:corpus-metadata}, the sampled repositories are highly visible, well-established, and collaboratively intensive.

\begin{table}[t]
\caption{Demographics of the sampled GitHub projects.}
\label{tab:corpus-metadata}
\centering
\footnotesize
\setlength{\tabcolsep}{3pt}

\begin{tabular}{@{}l rr c@{}}
\toprule
\textbf{Statistics} & \textbf{Mean} & \textbf{Median} & \textbf{Distribution} \\
\midrule
\# of Stars & 49,197.95 & 45,304.00 & \raisebox{-0.3\height}{\includegraphics[width=2.5cm,height=0.45cm]{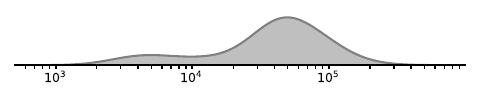}} \\
\# of Commits & 45,523.98 & 15,540.00 & \raisebox{-0.3\height}{\includegraphics[width=2.5cm,height=0.45cm]{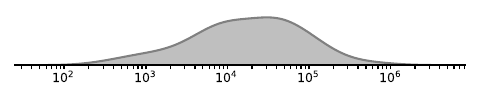}} \\
\# of Contributors & 1,666.12 & 1,045.50 & \raisebox{-0.3\height}{\includegraphics[width=2.5cm,height=0.45cm]{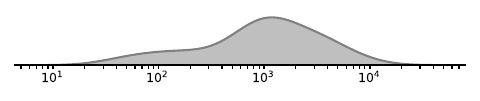}} \\
Age (years) & 9.86 & 10.03 & \raisebox{-0.3\height}{\includegraphics[width=2.5cm,height=0.45cm]{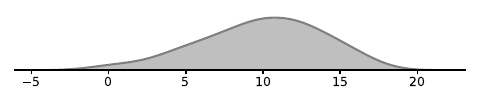}} \\
Policy adoption & 2025-07 & 2025-11 & \raisebox{-0.3\height}{\includegraphics[width=2.5cm,height=0.45cm]{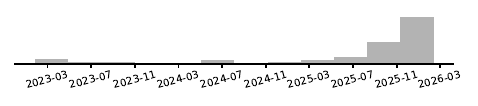}} \\
\midrule
Language & \multicolumn{3}{l}{Python (10), Go (8), C++ (8), JS (5), TS (4), others (20)} \\
\bottomrule
\end{tabular}
\end{table}

\subsection{Data Sources and Units of Analysis}

For each OSS project, we collected up to two kinds of public materials. The primary evidence consisted of \emph{repository-embedded governance materials}: README files, CONTRIBUTING files, GOVERNANCE files, SECURITY files, issue and PR templates, relevant files under \texttt{.github/}, standalone AI policy pages, and tool-facing files such as \texttt{AGENTS.md} or Copilot instructions.

The second kind of material consisted of \emph{surrounding justificatory materials}, such as maintainer-authored announcements, policy-related commit messages, linked issue or discussion threads, and project-linked external policy pages. Commits, issues, and PRs thus played two roles in our analysis: as snowballing paths when they introduced or modified GenAI rules (Section~3.2), and as context for interpreting the associated governance texts. In coding, these materials were treated as contextual evidence rather than standalone policy types: orientation assignments and strategy coding were anchored in stable governance text and only refined through justificatory materials when the two were consistent.

We analyzed the latest stable public version of each core governance text available as of February 2026 and used earlier discussions or commits to interpret policy formation and adoption context. For each project, we recorded the URL, access date, source type, governance surfaces, and the versioned location of the core policy text when available in a corpus sheet used as an audit trail.

Our analysis operated at two levels. At the project level, each OSS project formed a project-level policy case, consolidating all relevant documents from that project into a single unit, ensuring that the same stance expressed across multiple files was not counted as separate policies. At the text level, we coded strategy-bearing text segments: passages that expressed rules or judgments about admissibility, disclosure, responsibility, verification, provenance, reporting channels, sanctions, or infrastructural adjustment.

\subsection{Coding Procedure}

We analyzed the corpus through iterative qualitative coding with constant comparison, following a provisional coding approach~\cite{miles2014qualitative} in five steps: (1) drafting an initial codebook from community discussions and technical blog posts; (2) stabilizing it on an exploratory subset through independent coding by two authors; (3) independently coding the full corpus with the refined codebook; (4) computing inter-rater agreement; and (5) adjudicating remaining disagreements through joint re-review, with the third author resolving cases where the two coders still disagreed. The three stages below describe the analytical layers produced by this process.

\paragraph{Stage 1: open coding of maintainer concerns.}
Two authors independently conducted open coding on an initial heterogeneous subset of OSS projects to develop a shared codebook for recurring maintainer concerns. We then used the refined codebook to independently code the full corpus.

\paragraph{Stage 2: project-level memoing and governance orientations.}
After concern coding became stable, we wrote project-level memos summarizing each OSS project's main governance problem framing and relevant governance surfaces. On this basis, we assigned each OSS project a primary governance orientation by comparing its dominant governance logic across documents. Orientation was treated as a project-level interpretive abstraction rather than a document label. When a project displayed mixed signals, we re-reviewed the relevant texts jointly and assigned orientation based on the dominant governance judgment while retaining mixed or escalatory elements in the strategy analysis.

\paragraph{Stage 3: strategy abstraction and implementation analysis.}
We then compared coded OSS projects across the corpus to abstract cross-project governance strategies. We retained a strategy only when it was analytically distinct in governance function, behavioral target, and policy instruments. Throughout this stage, we repeatedly moved between project-level cases, coded segments, and emerging categories to refine strategy boundaries and clarify how strategies played different roles across governance orientations.

\label{sec:coding-procedure}
To strengthen analytic credibility, two authors independently performed both seed screening and full-corpus coding. Inter-rater agreement was computed at the case level after the codebook had stabilized (step~4) and before adjudication (step~5). For governance orientation, which assigned one dominant label to each of the 67 project-level cases, Cohen's kappa was 0.742. For concern and strategy coding, which allowed multiple codes per case, we calculated Cohen's kappa per code on binary presence/absence decisions and report the macro-average across codes: 0.745 for concerns and 0.840 for strategies. All concern and strategy counts reported below are case-level presence counts rather than mention frequencies, so repeated expressions within the same case were counted once. Most orientation disagreements clustered at the O2/O3 boundary. For the same OSS project, we compared multiple governance surfaces whenever possible and revisited borderline cases during adjudication. The full codebook is part of our replication package. For every concern, orientation, and strategy code, it gives an operationalized definition, rules on when to apply the code, and boundary notes distinguishing it from adjacent codes. The category descriptions in Sections~4 and~5 are condensed from these definitions.

\begingroup
\newcommand{\rqdisplayhead}[1]{\par\addvspace{7pt}\noindent\textbf{#1}\par\nobreak\smallskip\noindent}
\newcommand{\rqgrouphead}[1]{\par\addvspace{8pt}\noindent\textbf{#1}\par\nobreak\smallskip\noindent}
\newcommand{\rqruninhead}[1]{\par\addvspace{6pt}\noindent\textbf{#1}\ }

\section{RQ1: Maintainer Concerns}

Our corpus shows that GenAI enters OSS workflows well beyond code generation, including code contribution, discussion, issue triage, security reporting, and platform-level entry points. Maintainer concerns therefore span
seven recurring concern areas: 1) review bottlenecks and contributor responsibility in code contribution, 2) low-value AI text in communication,
3) rising triage costs at issue entry, 4) pressure on high-priority security
channels, 5) adversarial incentives, 6) copyright/licensing/provenance
uncertainty, and 7) platform defaults and tooling limits.

\rqruninhead{1. Code contribution.}
In code contribution, maintainers most consistently describe a \emph{review bottleneck}. GenAI makes PRs easier to generate without making them easier to review, explain, or validate. This concern appears in four closely related manifestations:

\begin{enumerate}[leftmargin=1.35em,label=(\arabic*),itemsep=2pt,topsep=3pt,parsep=0pt,partopsep=0pt]
\item \textbf{Intent and context.} A first recurring pressure is that AI makes it easier to submit changes whose scope outruns shared problem definition. Docusaurus warns that projects now receive ``1k LOC PRs that are obviously AI-generated and implement unsolicited features''~\cite{docusaurus_contributing_ai}. It also reminds contributors that significant changes require prior issue-based approval. The problem is not PR size alone, but that maintainers are pushed to decide during review whether the change should exist at all.
\item \textbf{Surface completeness and correctness.} A second recurring concern is that surface completeness no longer signals correctness. AI can produce polished, plausible-looking changes with test coverage that still fail at the semantic level. The Metasploit project described this pattern bluntly: some AI-assisted submissions were ``well-formatted, looked really neat, and did nothing it said it did''~\cite{metasploit_contributing_ai}. Review therefore cannot shift from correctness checking to style checking.
\item \textbf{Authorship and responsibility.} A third concern is whether a contributor actually understands, explains, and owns the change. typescript-eslint captures this especially sharply: maintainers do not want review to become ``effectively babysitting someone's Claude instance via the code review process''~\cite{typescript_eslint_issue_11416}. The issue is not AI use alone, but whether a human author remains accountable.
\item \textbf{Review capacity.} A fourth concern is that review capacity becomes the first bottleneck. Projects therefore describe AI contribution primarily as a review problem rather than a generation problem. Oh My Zsh summarizes the asymmetry succinctly: ``review is the bottleneck. Not code generation.''~\cite{ohmyzsh_humans_loop}. FastAPI makes the same point even more forcefully, describing low-effort AI submissions as ``a Denial-of-service attack on our human effort''~\cite{fastapi_automated_code_ai}.
\end{enumerate}

\rqruninhead{2. Communication.}
In communication and discussion, the same responsibility problem appears as \emph{low-value AI text} and weaker human exchange.
typescript-eslint states this directly: AI-generated PR descriptions are often ``nearly-verbatim descriptions of the code diff, which add no value to the PR''~\cite{typescript_eslint_ai_policy}. Jaeger states the corresponding expectation for review interaction just as succinctly: ``Code review is a discussion between people, not bots''~\cite{jaeger_ai_usage_policy}. The objection is therefore not simply to verbosity, but to generated text that adds little context and displaces direct human exchange.

\rqruninhead{3. Issue triage.}
At issue entry, the problem is \emph{triage costs}: AI makes reports look plausible without making them more actionable.
NewPipe encodes this problem explicitly by banning AI-written issue and PR templates, stating that ``Those texts are often lengthy and miss critical information''~\cite{newpipe_ai_policy}. What matters at this stage is whether a report contains the specific information needed for triage.

\rqruninhead{4. Security reporting.}
Security reporting is the same triage problem in \emph{high-priority channels}.
curl states this concern particularly clearly: because security reports are investigated with priority, ``fake and otherwise made up security problems effectively prevent us from doing real project work and make us waste time and resources''~\cite{curl_ai_use}. Fabricated or unverifiable AI reports are especially costly because they immediately consume scarce security-handling time.

\rqruninhead{5. Adversarial incentives.}
In some contexts, maintainers frame the problem as adversarial incentives. When rewards remain in place, AI lowers the cost of producing bad-faith or low-value reports.
curl provides the clearest example in the context of bug bounties: ``A bug bounty gives people too strong incentives to find and make up `problems' in bad faith that cause overload and abuse''~\cite{curl_bug_bounty_end}. Here the concern is not mere noise, but noise made worthwhile by institutional incentives.

\rqruninhead{6. Provenance and licensing.}
For some projects, the concern is not output quality but copyright/licensing/provenance \emph{uncertainty}.
QEMU states the legal uncertainty directly: with AI content generators, ``the copyright and license status of the output is ill-defined with no generally accepted, settled legal foundation''~\cite{qemu_code_provenance_ai}. NetBSD operationalizes the same concern even more bluntly by treating LLM output as ``presumed to be tainted code''~\cite{netbsd_commit_guidelines_ai}. Here the issue is not ordinary review quality, but whether such inputs can be admitted at all.

\rqruninhead{7. Platform and tooling.}
Finally, some projects locate part of the problem in \emph{platform defaults/tooling limits} that amplify the concerns above.
Zig's maintainers describe GitHub's Copilot issue composer as taking a user's ``simple explanation of a bug'' and rewriting it into ``a multi-paragraph monologue with no interesting details and added false information''~\cite{zig_issue_warning}. tldraw links its temporary closure of external PRs directly to infrastructure limits, calling it ``a temporary policy until GitHub provides better tools for managing contributions''~\cite{tldraw_contributions_policy}. Here the issue is not only what contributors do, but what platforms make easy and what projects still cannot control.

\section{RQ2: Governance Orientations and Strategies}

To answer RQ2, we first identify three governance orientations that explain why projects facing similar GenAI pressures develop markedly different institutional responses. We then derive 12 concrete governance strategies through which these orientations are operationalized across collaboration surfaces.

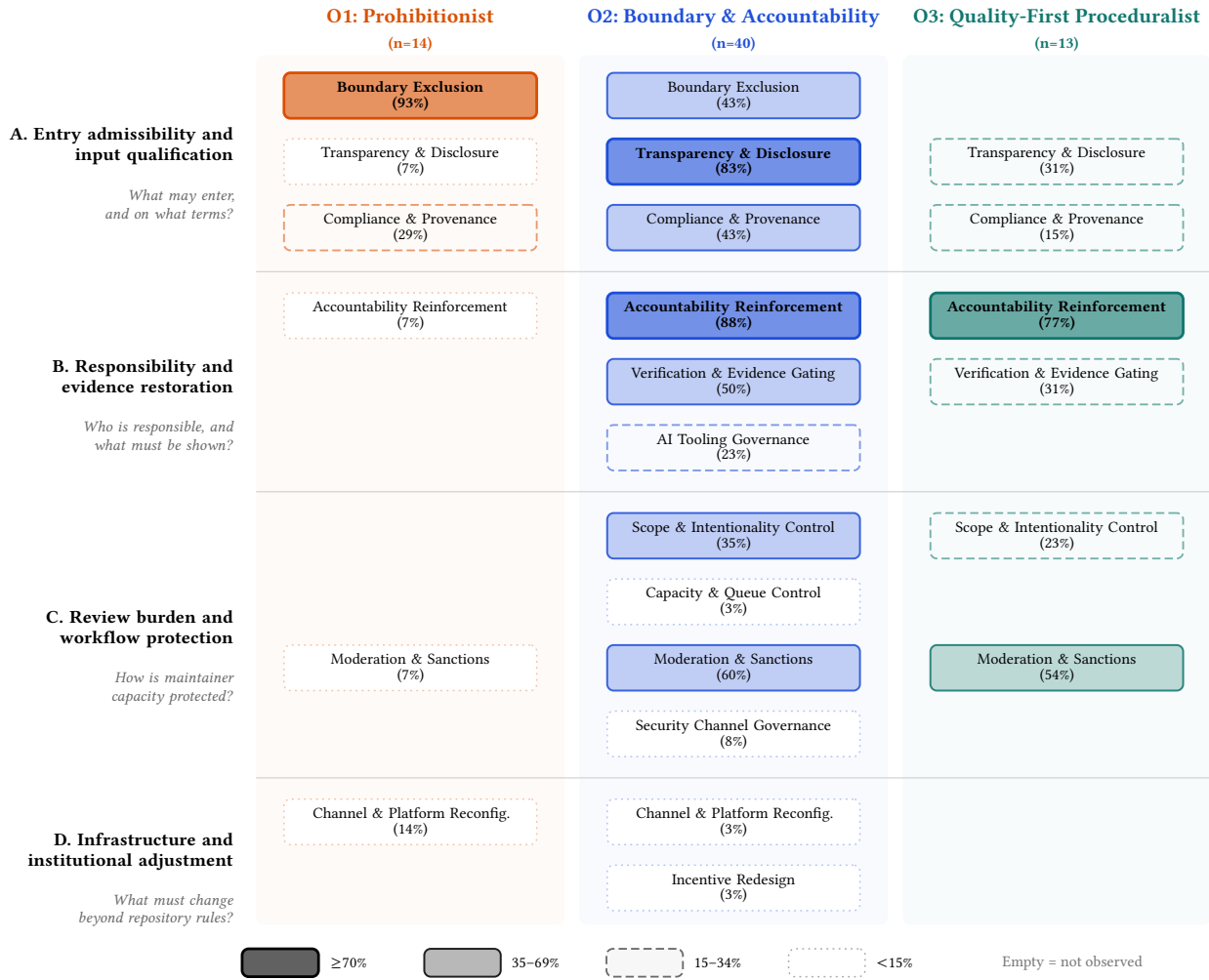
\begin{figure*}[t]
\centering
\newcommand{\countline}[1]{\\[-1pt]{\fontsize{6}{6.8}\selectfont (#1\%)}}
\begin{tikzpicture}[
    svhigh/.style={
        rectangle, rounded corners=2.5pt, draw=#1!100, fill=#1!62,
        line width=1.05pt, minimum height=0.62cm, text width=3.3cm,
        align=center, font=\scriptsize\bfseries, inner sep=2pt,
    },
    shigh/.style={
        rectangle, rounded corners=2.5pt, draw=#1!88, fill=#1!28,
        line width=0.75pt, minimum height=0.62cm, text width=3.3cm,
        align=center, font=\scriptsize, inner sep=2pt,
    },
    smed/.style={
        rectangle, rounded corners=2.5pt, draw=#1!58, fill=#1!3,
        dash pattern=on 3.2pt off 1.6pt,
        line width=0.7pt, minimum height=0.62cm, text width=3.3cm,
        align=center, font=\scriptsize, inner sep=2pt,
    },
    slow/.style={
        rectangle, rounded corners=2.5pt, draw=#1!34, fill=white,
        dash pattern=on 0.7pt off 1.65pt,
        line width=0.55pt, minimum height=0.62cm, text width=3.3cm,
        align=center, font=\scriptsize, inner sep=2pt,
    },
]

\definecolor{o1c}{HTML}{D94F00}
\definecolor{o2c}{HTML}{1D4ED8}
\definecolor{o3c}{HTML}{0F766E}

\def\colW{4.2}
\def\colGap{0.2}
\def\rowH{1.0}
\def\labelW{3.32}
\def\headerH{1.1}

\pgfmathsetmacro{\xA}{\labelW + \colW/2}
\pgfmathsetmacro{\xB}{\labelW + \colW + \colGap + \colW/2}
\pgfmathsetmacro{\xC}{\labelW + 2*\colW + 2*\colGap + \colW/2}
\pgfmathsetmacro{\totalW}{\labelW + 3*\colW + 2*\colGap}

\node[font=\small\bfseries, text=o1c, align=center] at (\xA, 0)
    {O1: Prohibitionist\\{\scriptsize (n=14)}};
\node[font=\small\bfseries, text=o2c, align=center] at (\xB, 0)
    {O2: Boundary \& Accountability\\{\scriptsize (n=40)}};
\node[font=\small\bfseries, text=o3c, align=center] at (\xC, 0)
    {O3: Quality-First / Tool-Agnostic\\{\scriptsize (n=13)}};

\begin{scope}[on background layer]
  \fill[o1c!3, rounded corners=3pt]
    (\labelW, -0.35) rectangle (\labelW+\colW, -12.2);
  \fill[o2c!3, rounded corners=3pt]
    (\labelW+\colW+\colGap, -0.35) rectangle (\labelW+2*\colW+\colGap, -12.2);
  \fill[o3c!3, rounded corners=3pt]
    (\labelW+2*\colW+2*\colGap, -0.35) rectangle (\labelW+3*\colW+2*\colGap, -12.2);
\end{scope}

\node[font=\footnotesize\bfseries, text width=3.05cm, align=right, anchor=east]
    at (\labelW-0.2, -1.58) {A. Entry admissibility and\\input qualification};
\node[font=\scriptsize, text width=3.05cm, align=right, anchor=east, text=black!60]
    at (\labelW-0.2, -2.38) {\textit{What may enter,\\and on what terms?}};

\node[font=\footnotesize\bfseries, text width=3.05cm, align=right, anchor=east]
    at (\labelW-0.2, -4.72) {B. Responsibility and\\evidence restoration};
\node[font=\scriptsize, text width=3.05cm, align=right, anchor=east, text=black!60]
    at (\labelW-0.2, -5.53) {\textit{Who is responsible, and\\what must be shown?}};

\node[font=\footnotesize\bfseries, text width=3.05cm, align=right, anchor=east]
    at (\labelW-0.2, -8.17) {C. Review burden and\\workflow protection};
\node[font=\scriptsize, text width=3.05cm, align=right, anchor=east, text=black!60]
    at (\labelW-0.2, -8.98) {\textit{How is maintainer\\capacity protected?}};

\node[font=\footnotesize\bfseries, text width=3.05cm, align=right, anchor=east]
    at (\labelW-0.2, -11.2) {D. Infrastructure and\\institutional adjustment};
\node[font=\scriptsize, text width=3.05cm, align=right, anchor=east, text=black!60]
    at (\labelW-0.2, -12.0) {\textit{What must change\\beyond repository rules?}};

\draw[black!20, line width=0.4pt]
    (\labelW, -3.3) -- (\totalW, -3.3);
\draw[black!20, line width=0.4pt]
    (\labelW, -6.3) -- (\totalW, -6.3);
\draw[black!20, line width=0.4pt]
    (\labelW, -10.2) -- (\totalW, -10.2);

\node[svhigh=o1c] at (\xA, -0.9)
    {Boundary Exclusion\countline{93}};
\node[shigh=o2c] at (\xB, -0.9)
    {Boundary Exclusion\countline{43}};

\node[slow=o1c] at (\xA, -1.8)
    {Transparency \& Disclosure\countline{7}};
\node[svhigh=o2c] at (\xB, -1.8)
    {Transparency \& Disclosure\countline{83}};
\node[smed=o3c] at (\xC, -1.8)
    {Transparency \& Disclosure\countline{31}};

\node[smed=o1c] at (\xA, -2.7)
    {Compliance \& Provenance\countline{29}};
\node[shigh=o2c] at (\xB, -2.7)
    {Compliance \& Provenance\countline{43}};
\node[smed=o3c] at (\xC, -2.7)
    {Compliance \& Provenance\countline{15}};

\node[slow=o1c] at (\xA, -3.9)
    {Accountability Reinforcement\countline{7}};
\node[svhigh=o2c] at (\xB, -3.9)
    {Accountability Reinforcement\countline{88}};
\node[svhigh=o3c] at (\xC, -3.9)
    {Accountability Reinforcement\countline{77}};

\node[shigh=o2c] at (\xB, -4.8)
    {Verification \& Evidence Gating\countline{50}};
\node[smed=o3c] at (\xC, -4.8)
    {Verification \& Evidence Gating\countline{31}};

\node[smed=o2c] at (\xB, -5.7)
    {AI Tooling Governance\countline{23}};

\node[shigh=o2c] at (\xB, -6.9)
    {Scope \& Intentionality Control\countline{35}};
\node[smed=o3c] at (\xC, -6.9)
    {Scope \& Intentionality Control\countline{23}};

\node[slow=o2c] at (\xB, -7.8)
    {Capacity \& Queue Control\countline{3}};

\node[slow=o1c] at (\xA, -8.7)
    {Moderation \& Sanctions\countline{7}};
\node[shigh=o2c] at (\xB, -8.7)
    {Moderation \& Sanctions\countline{60}};
\node[shigh=o3c] at (\xC, -8.7)
    {Moderation \& Sanctions\countline{54}};

\node[slow=o2c] at (\xB, -9.6)
    {Security Channel Governance\countline{8}};

\node[slow=o1c] at (\xA, -10.8)
    {Channel \& Platform Reconfig.\countline{14}};
\node[slow=o2c] at (\xB, -10.8)
    {Channel \& Platform Reconfig.\countline{3}};

\node[slow=o2c] at (\xB, -11.7)
    {Incentive Redesign\countline{3}};

\pgfmathsetmacro{\legOne}{3.65}
\pgfmathsetmacro{\legStep}{2.48}
\foreach \style/\label/\idx in {
    svhigh/{$\geq$70\%}/0,
    shigh/{35--69\%}/1,
    smed/{15--34\%}/2,
    slow/{$<$15\%}/3
} {
    \pgfmathsetmacro{\legX}{\legOne + \idx*\legStep}
    \node[\style=black, text width=0.9cm, minimum height=0.38cm, font=\scriptsize]
        at (\legX, -12.72) {};
    \node[font=\scriptsize, anchor=west]
        at (\legX + 0.56, -12.72) {\label};
}
\node[font=\scriptsize, anchor=west, text=black!55]
    at (\legOne + 4*\legStep - 0.2, -12.72) {Empty = not observed};

\end{tikzpicture}
\caption{Governance strategies mapped by functional group (rows) and governance orientation (columns). Shading intensity reflects within-orientation prevalence using four bands ($\geq$70\%, 35--69\%, 15--34\%, and $<$15\%). Percentages in parentheses show the share of cases within that orientation that exhibit the strategy, while orientation totals are reported in the column headers. Empty cells indicate that the strategy was not observed in that orientation.}
\label{fig:strategy-orientation}
\end{figure*}

\subsection{Governance orientations}

After open-coding maintainer concerns, we compared how projects interpreted these problems, where they located core risks, and how they allocated governance responsibility. We found that cross-project variation cannot be explained simply in terms of whether a project ``allows AI.'' Instead, through iterative comparison, we derived three recurring governance orientations---higher-order judgments, not static labels or maturity rankings---that reflect three distinct ways of framing the governance problem: whether certain inputs should enter the workflow at all; whether they may enter under explicit boundaries of accountability, visibility, and verification; or whether they should be absorbed into existing quality and maintainer-cost controls. Organizational form and domain context function as boundary conditions that shape where pressure first arises, but do not reliably predict which orientation a project adopts.
Across the 67 cases, O2 was the most common dominant orientation (40 cases, 59.7\%), followed by O1 (14, 20.9\%) and O3 (13, 19.4\%). Twenty-three cases displayed mixed governance logics, combining a dominant orientation with material elements of a secondary one.

\rqruninhead{O1: Prohibitionist / zero-tolerance.}
The core judgment behind O1 is that certain GenAI-related inputs present \emph{structural, non-absorbable risk}---including provenance uncertainty, copyright and licensing incompatibility, or taintedness---and should therefore \emph{not} be absorbed into the existing review process. The prior question is not how to review such inputs more effectively, but whether they should enter the workflow at all. QEMU states that it will ``DECLINE any contributions which are believed to include or derive from AI generated content''~\cite{qemu_code_provenance_ai}. NetBSD similarly states that LLM-generated code ``is presumed to be tainted code, and must not be committed without prior written approval by core''~\cite{netbsd_commit_guidelines_ai}. These texts define AI-generated code as a class of input whose admissibility is in question from the outset.

\rqruninhead{O2: Boundary-and-accountability.}
The core judgment behind O2 is that GenAI-related inputs may enter the workflow, but only under explicit conditions of accountability, visibility, and verification. Rather than excluding AI-mediated input categorically, this orientation treats AI as a distinct governance object requiring explicit rules across multiple collaboration surfaces (\emph{multi-surface boundary governance}). CloudNativePG states that the goal is ``not to ban AI entirely,'' but to shift ``the burden of proof back to the contributor,'' requiring ``full human accountability for any AI-assisted work''~\cite{cloudnativepg_ai_policy_issue}. llama.cpp draws the boundary more explicitly: AI tools may be used ``solely in an assistive capacity,'' and ``all AI usage requires explicit disclosure''~\cite{llamacpp_ai_policy}. O2 thus locates the governance problem in AI-mediated input itself and admits such input only under AI-specific rules of disclosure, human accountability, verification, or enforcement.

\rqruninhead{O3: Quality-first / tool-agnostic.}
The core judgment behind O3 is that projects prefer to absorb AI-related inputs into existing expectations around contribution quality, maintainability, and maintainer cost, rather than building AI-specific governance rules (\emph{reasserting existing quality thresholds under cheaper-generation conditions}). The key question is not whether AI was used, but whether the input is reviewable, maintainable, and worth maintainers' limited time. Oh My Zsh expresses this clearly: ``Not a moral stance. Not a ban. Just clarity around expectations and accountability'' and ``We're not interested in policing tooling. We're interested in accountability.''~\cite{ohmyzsh_humans_loop}. curl articulates the same logic: licensing and quality obligations apply ``independent if AI is used or not''~\cite{curl_ai_use}. O3 thus locates the problem not in AI use per se, but in ordinary contribution quality under cheaper-generation conditions, addressed through general controls of reviewability, testing, scope, and maintainability.

\rqruninhead{Boundary conditions.}\label{sec:boundary-conditions}
The boundaries among the three orientations lie in higher-order judgments about what kind of governance problem GenAI presents, not in any single device such as disclosure or templates. The O2/O3 boundary was the least sharp. Our tie-breaker was whether a project introduced AI-specific admissibility or accountability boundaries (O2), or instead reasserted quality-first review controls that apply regardless of AI use (O3). The llama.cpp and curl policies quoted above illustrate the difference: llama.cpp admits AI use only under mandatory disclosure, whereas curl applies the same obligations whether or not AI is used. Orientations are dominant case-level logics, and strategies can co-occur with any orientation. Under O1, for example, disclosure supports narrow exceptions or enforcement rather than contradicting prohibition. The three orientations should therefore be read as ideal-typical governance logics rather than static labels or a ranking of policy strictness.

\subsection{Governance strategies}

While governance orientations capture \emph{why} projects respond differently, governance strategies capture \emph{how} those orientations are operationalized across contribution workflows. Through cross-project comparison, we derived 12 strategies that cluster into four functional groups: \emph{entry admissibility and input qualification}, \emph{responsibility and evidence restoration}, \emph{review burden and workflow protection}, and \emph{infrastructure and institutional adjustment}. Each strategy takes effect through \emph{policy instruments}, the textual and workflow mechanisms a project deploys. Typical examples include disclosure checkboxes and template fields, evidence requirements, PR limits, security-channel rules, tool-facing instruction files, and platform configuration changes. Figure~\ref{fig:strategy-orientation} maps their \emph{within-orientation prevalence}. In broad terms, O1 concentrates on front-gate and infrastructural controls, O2 carries the broadest AI-specific governance layer, and O3 relies more heavily on general-purpose quality controls while adopting fewer explicit AI-specific mechanisms.

\rqgrouphead{Function A. Entry admissibility and input qualification.}

These strategies move governance toward the point of intake, addressing what kinds of inputs may enter the system and what qualifying information must accompany them.

\rqdisplayhead{A1. Boundary Exclusion / Reject-or-Ban.}
Boundary exclusion governs the behavior of \emph{excluding certain classes of GenAI-mediated input from the acceptable contribution surface altogether}. Unlike disclosure, accountability, or verification, this strategy does not ask how to review GenAI-related input more effectively. It asks a prior admissibility question: whether such input should be treated as reviewable at all. Accordingly, its clearest expressions are not requests for explanation, but statements such as ``will not accept,'' ``does not allow,'' ``decline,'' or ``must not be committed.''

As illustrated by QEMU's and NetBSD's explicit exclusion language (see O1 above), boundary exclusion operates not as a stricter review rule, but as a refusal to treat certain inputs as ordinarily admissible.

The main benefit is front-loading the forms of uncertainty maintainers most want to avoid at the stage of admissibility judgment, reducing the need to repeatedly re-litigate boundaries during review. The cost is equally clear: projects narrow the surface of acceptable participation. Boundary exclusion is near-universal in O1 (93\%) but also appears in 43\% of O2 projects, where it typically applies selectively to specific contribution surfaces---such as fully generated PRs or unverified security reports---rather than as a blanket prohibition.

\rqdisplayhead{A2. Transparency and Disclosure.}
Disclosure seeks not to determine whether AI may be used, but to turn AI use from a hidden practice into a visible, discussable, and reviewable input signal. Its immediate target is contributors' reporting behavior when submitting issues, PRs, comments, or reports. What maintainers need is not abstract transparency for its own sake, but a way to calibrate expectations once inputs become longer, more polished, and more contribution-like. Disclosure helps them distinguish between what appears to come from the contributor's own understanding and what depends heavily on model-generated output.

Spec Kit states that if a contributor uses ``any kind of AI assistance,'' it ``must be disclosed in the pull request or issue,'' and adds that failure to disclose makes it harder ``to determine how much scrutiny to apply''~\cite{spec_kit_ai_contributions}. This captures disclosure not as moral signaling, but as review calibration. In more formalized settings, the same logic is pushed into templates or dedicated fields so that AI use becomes intake information rather than an after-the-fact suspicion. A few projects additionally request documentation of prompts or AI interactions. We observed this too rarely for a standalone strategy, but it works as a policy instrument supporting disclosure, provenance, or evidence gating.

The main benefit of disclosure is that it reduces maintainers' up-front information-discovery cost. They do not need to first infer how an input was produced before deciding how to process it. At the same time, the trade-off is concrete: contributors must bear additional metadata-reporting costs, and explicit AI marking may generate concerns about differential treatment. For this reason, many projects pair disclosure requirements with explicit statements that they do not intend to police tools per se, but to use disclosure as a more rational review-calibration signal.

\rqdisplayhead{A3. Compliance and Provenance Safeguarding.}
Unlike boundary exclusion, compliance and provenance safeguards do not necessarily require direct prohibition. More often, they appear in projects that allow some degree of AI use while insisting that contributors retain responsibility for redistributability, license compatibility, and lawful provenance. The central behavior being governed is therefore the contributor's obligation to make these judgments before submission, rather than leaving them for maintainers to reconstruct case by case during review.

mpv states that AI-assisted contributions ``are not forbidden,'' but the submitter ``takes full responsibility'' and must confirm that the code can be submitted under the relevant license; curl similarly insists that ``the burden is on them to ensure no unlicensed code is submitted''~\cite{mpv_ai_assisted_contributions,curl_ai_use}. The core move is to push provenance and redistributability checks back onto contributors before technical review begins.

The benefit is reallocating legal and provenance uncertainty away from maintainers. The cost falls on contributors, who must evaluate not only whether the content works but whether it can be redistributed under clear provenance. This strategy's role varies sharply by orientation: in O1, provenance uncertainty serves as a rationale for categorical exclusion; in O2, it becomes a contributor-facing compliance obligation; in O3, it is folded into general licensing responsibility.

\rqgrouphead{Function B. Responsibility and evidence restoration.}

These strategies address how projects restore accountability, explanation, and evidence where GenAI makes them thinner.

\rqdisplayhead{B1. Accountability Reinforcement.}
Accountability reinforcement seeks to pull responsibility back from the tool to the named human contributor. It does not primarily regulate whether AI was used. Instead, it regulates whether the submitter understands the content, can explain it during review, and is willing to remain responsible for revising, defending, and maintaining it. Accountability indicates answerability and ownership in review, not blame. In this way, the strategy redefines authorship as a responsibility relation rather than a mere act of submission.

CloudNativePG states that ``The human contributor is the sole party responsible for the contribution,'' and that ``The AI generated it'' is ``never an acceptable answer''; Jaeger echoes the same expectation with ``You own everything you submit'' and ``Code review is a discussion between people, not bots''~\cite{cloudnativepg_ai_policy,jaeger_ai_usage_policy}. These policies treat authorship less as the act of opening a PR than as the continuing duty to explain, revise, and stand behind it.

The principal benefit is reducing maintainers' uncertainty about who really understands the change and will remain responsible for it. The trade-off is substantial: this strategy raises the bar for participation by tying admissibility to explainability and sustained responsibility. In O2, accountability reinforcement forms part of an explicit AI-specific boundary regime (``The AI generated it is never an acceptable answer''), whereas in O3 it is more often absorbed into general contribution-quality expectations.

\rqdisplayhead{B2. Verification and Evidence Gating.}
Verification and evidence gating seeks to change contributor behavior by requiring sufficient supporting material before an input is processed further. This may include local test logs, proof-of-concept artifacts, reproduction steps, references, or explicit human validation notes. The key governance move is not to ask reviewers to be more careful, but to shift the burden of demonstrating verifiability toward the submitter.

Kornia encodes verification as an explicit gate: ``Every PR introducing functional changes must include a pasted snippet of the local test logs,'' and implementations must cite an ``existing library reference'' for verification~\cite{kornia_ai_policy}. In security reporting, similar hard evidence thresholds apply (see Security Channel Governance below). These rules shift the burden from reviewer inference to submitter-provided evidence.

The benefit is reducing secondary verification cost during review; the trade-off is a clear increase in submission preparation cost. Verification is more prevalent in O2 (50\%) than O3 (31\%), reflecting O2's tendency to add evidence requirements as part of its AI-specific boundary regime, whereas O3 projects more often rely on pre-existing quality gates without adding new AI-motivated evidence thresholds.

\rqdisplayhead{B3. AI Tooling Governance.}
AI tooling governance belongs in this group because it also seeks to restore responsibility and control, but does so at the tool layer rather than only at the human layer. Compared with accountability and verification, it is a further step: projects do not merely regulate how humans should use AI, but begin to regulate AI tools and agents directly through files such as \texttt{AGENTS.md}, Copilot instructions, or other tool-facing repository rules.

JabRef's \texttt{AGENTS.md} makes the tool-facing nature of this strategy explicit: agents must ``Never commit generated code without human review,'' must not ``Write entire PRs,'' and must not ``Automate the submission of code changes''~\cite{jabref_agents_policy}. The result is governance that addresses not only human conduct, but the expected behavior of the tools themselves.

The benefit of tooling governance is that it allows some governance work to happen \emph{before} low-quality or poorly attributable inputs reach review, by constraining the tool's role directly. It therefore shifts some control from ex post evaluation to ex ante interface design. Its costs are also distinctive. First, it depends heavily on whether tools and agent ecosystems actually read and comply with repository-level instructions. Second, projects themselves must maintain a layer of policy text written for machine consumption and keep it aligned with ordinary contributor-facing rules. The challenge here is therefore not only rule design, but the stability of the interface between repository policy and tool behavior.

\rqgrouphead{Function C. Review burden and workflow protection.}

These strategies protect maintainer time, review queues, and high-priority channels once inputs have already entered the system.

\rqdisplayhead{C1. Scope and Intentionality Control.}
Scope and intentionality control governs not only PR size, but whether a contribution is framed within the right problem context. Projects use this strategy to require small, focused, issue-linked, and intentionally scoped submissions, thereby discouraging broad, speculative, or misaligned PRs that arrive before the underlying problem has been accepted.

CloudNativePG states ``Design First'' for non-trivial changes and warns that PRs arriving ``out of the blue'' may be closed if they do not align with project context~\cite{cloudnativepg_ai_policy}. PyTorch formalizes the same logic upstream by stating that ``PRs must have an associated actionable Issue''~\cite{pytorch_ai_assisted_development}. In both cases, the point is to ensure that review does not become the place where maintainers must first reconstruct why the work should exist.

The benefit of this strategy is that it lowers the cost of problem redefinition during review. Maintainers are less often forced to ask why a contribution exists, why it takes its current scope, and whether it fits the project's priorities. The trade-off is that more filtering work moves upstream into issue triage and proposal stages, reducing the space for drive-by contributions and fast exploratory participation.

\rqdisplayhead{C2. Capacity and Queue Control.}
Capacity and queue control treats the review queue itself as a scarce resource in need of governance. Instead of focusing on the technical content of an individual submission, it regulates input rate, concurrency, and review load. This strategy makes explicit a practical reality of OSS maintenance: even when contributors act in good faith, review bandwidth remains limited, and GenAI can significantly increase the rate at which reviewable-looking inputs are produced.

Jaeger states that ``we limit the number of simultaneous open PRs for new contributors'' and that excess PRs may be labeled ``on-hold'' or closed~\cite{jaeger_pr_limits}. Here the governance target is not the correctness of any single patch, but how much simultaneous reviewer attention a contributor may occupy. The benefit is making maintainer time governable as a resource; the cost is that waiting time and participation friction are reallocated to contributors, especially newcomers.

\rqdisplayhead{C3. Moderation and Sanctions.}
Moderation and sanctions govern what happens once certain classes of input have already been judged low-effort, repeatedly non-compliant, adversarial, or not worth further processing. Rather than trying to improve such inputs, this strategy defines more consistent ways to close, reject, block, or ban them, thereby reducing the cost of repeated case-by-case negotiation.

pandas states that maintainers may ``ban users from contributing'' if they violate the guidelines repeatedly; CPython likewise notes that unproductive AI-generated issues or PRs may be closed and repeat submitters ``may be blocked''; curl goes further in security contexts: ``We ban users immediately who submit made up fake reports''~\cite{pandas_automated_contributions_policy,cpython_generative_ai,curl_ai_use}. Moderation is widespread in both O2 and O3 but functions less as a standalone governance logic than as an enforcement amplifier whose effectiveness depends on other strategies---disclosure, verification, or scope control---already being in place.

\rqdisplayhead{C4. Security Channel Governance.}
Security channel governance belongs in this group because it protects a particularly scarce and high-value workflow: vulnerability handling. Unlike ordinary PRs or general issue reports, security submissions often receive elevated processing priority. Security channel governance also differs from the general evidence gating of B2. B2 gates ordinary PR and issue handling, whereas C4 protects high-priority security channels, where fake reports (fabricated, hallucinated, or unverifiable vulnerability reports) immediately consume scarce incident-response time. This strategy therefore governs what counts as an actionable security input and what level of evidence, disclosure, and reproducibility is required before high-priority maintainer resources are engaged.

curl warns that it is ``rarely a good idea to just copy and paste an AI generated report'' and instead tells reporters to ``write the report yourself'' after verifying the problem~\cite{curl_ai_use}. llama.cpp adds a hard evidence threshold by requiring that ``Your report must include a working Proof-of-Concept''~\cite{llamacpp_security_policy}. The shared logic is to front-load actionability before scarce security attention is engaged.

The benefit of this strategy is that it prevents fabricated, speculative, or unverifiable inputs from crowding out real vulnerabilities. The trade-off is a higher reporting threshold: weaker but potentially useful leads may be filtered out if they do not meet the required standard of evidence. Projects thus exchange lower waste in security triage for a narrower and more demanding reporting interface.

\rqgrouphead{Function D. Infrastructure and institutional adjustment.}

When repository-level rules are insufficient, some projects adjust the surrounding infrastructure or institutional conditions under which contributions enter the system.

\rqdisplayhead{D1. Channel, Venue, and Platform Reconfiguration.}
Channel, venue, and platform reconfiguration governs not primarily \emph{what} contributors submit, but \emph{where} and \emph{through which interface} they are allowed to submit it. When maintainers conclude that repository-level rules can no longer control the relevant problems, they may suspend external PRs, redirect reporting channels, or even migrate their main collaboration venue.

tldraw states that it is ``not accepting pull requests from external contributors'' and that such PRs ``will be automatically closed'' until GitHub provides better tools for managing contributions~\cite{tldraw_contributions_policy}. Zig represents a stronger infrastructural move: after migrating to Codeberg, it explicitly expected ``fewer violations'' of its strict no-AI policy because GitHub had been pushing AI-assisted entry points so aggressively~\cite{zig_migrating_codeberg}. Here the governance response is not to review inputs differently, but to change the channel through which they arrive.

The benefit of this strategy is that it changes the entry architecture itself rather than asking maintainers to repeatedly interpret and enforce the same boundaries after the fact. Its costs, however, are substantial. Contributors must adapt to new channels or platforms, and projects assume new infrastructural burdens and potential ecosystem friction. For this reason, reconfiguration typically appears as an escalation strategy when repository-level governance is judged insufficient.

\rqdisplayhead{D2. Incentive Redesign.}
Incentive redesign is one of the least frequent but most revealing strategies in the corpus. It does not regulate how a particular input is written or reviewed. Instead, it targets the broader reward structure that makes certain kinds of harmful input worth producing in the first place. In this sense, it governs not content quality directly, but the institutional conditions that generate persistent low-value or adversarial submissions.

curl makes this logic unusually explicit: it removed rewards ``to remove the incentives for submitting made up lies'' and later generalized that ``A bug bounty gives people too strong incentives to find and make up problems in bad faith''~\cite{curl_bug_bounty_end}. By addressing harmful input at the level of institutional reward rather than downstream filtering, this strategy functions as a high-cost governance escalation that may also discourage some good-faith participation.

\endgroup

\section{Threats to Validity and Limitations}

\subsection{Coverage and Transferability}

Our corpus intentionally privileges visible OSS projects and policy-rich governance environments. The seed frame began with the top-ranked repositories on Gitstar Ranking's repository leaderboard and was then extended through theoretically motivated snowball sampling, resulting in a varied and reasonably sized corpus. Accordingly, the findings should be interpreted as an analytically transferable account of publicly visible GenAI governance rather than a prevalence estimate for the entire OSS ecosystem. Smaller projects, low-activity repositories, non-GitHub communities, and non-English communities may encode similar pressures less explicitly or govern them differently.

\subsection{Publicly Encoded Governance vs.\ Enacted Practice}

This study analyzes governance as it is publicly expressed and institutionally encoded in project materials. Actual governance may also depend on private maintainer discussions, discretionary triage, informal coordination, moderation practices, or security workflows that are not visible in repository texts. Our findings should therefore be read as an account of public governance interfaces rather than a complete account of every path through which OSS projects actually process AI-related contributions. We also do not directly evaluate the causal effectiveness of these rules or whether they are enforced uniformly. Our results are coded from public texts and have not been validated by the repository owners or maintainers, so our interpretations of individual policies may contain imprecisions. Most analyzed materials are maintainer-authored or directly linked from the projects' own governance texts, so this limitation is unlikely to invalidate the descriptive taxonomy, though it does bound what we can claim about intent, enforcement, and outcomes. Maintainer interviews and surveys remain important future work.

\subsection{Interpretive and Temporal Limits}

These categories remain interpretive abstractions rather than naturally given classes. The three orientations are best understood as ideal-typical dominant logics, and orientation counts should therefore be read as approximate patterning rather than sharp natural groupings, especially near the O2/O3 boundary. Strategy findings are less sensitive to that boundary because strategy presence was coded independently of orientation.

In addition, GenAI-related governance is changing quickly: platforms, tools, and project rules continue to shift over short periods of time. Our analysis is thus a cross-sectional account of one time window, and later governance changes may alter how particular OSS projects are best interpreted.

\section{Discussion}

\subsection{GenAI Reshapes and Front-loads OSS Governance}
Our results demonstrate that GenAI governance is rarely a from-scratch invention.
Instead, it represents a front-loading redesign of existing governance mechanisms.
Pull-based development has always faced ``review friction''~\cite{gousios2014exploratory}, and the open source communities have invented strategies, including contribution and security guidelines, issue and PR templates, proposal and approval workflows, to gate external contributions, balance maintainer workload, and coordinate collaboration efforts.

GenAI creates a radical socio-technical asymmetry, where
the cost of producing a plausible, complete-looking PR approaches zero, while the cost of reviewing it remains high or even increases.
Consequently, we observe a shift from \emph{downstream code review} to \emph{upstream admission control}.
Strategies like \emph{Scope and Intentionality Control} (requiring pre-approved issues) or \emph{Verification and Evidence Gating} (requiring proof-of-concept before review) are attempts to protect the most scarce resource in OSS: maintainer attention~\cite{eghbal2020working}.
By forcing contributors to justify the ``why'' and ``how'' before maintainers look at the ``what,'' projects are effectively moving the triage line further up the workflow to survive the ``denial-of-service attack'' of AI-assisted inputs.

\subsection{From Repository to Infrastructure-Level Governance}

While most projects in the corpus have shown agility in updating governance policies in markdown files, our results reveal the limits of repository-level governance. Text-based rules (CONTRIBUTING, SECURITY) are often reactive and might easily be bypassed by coding agents or low-effort human submitters.
The platform-and-tooling concern identified in RQ1 suggests that as long as platforms provide a ``one-click'' AI generation path without corresponding identification or intake controls, repository rules will remain under siege.

We are beginning to see the emergence of infrastructure-level countermeasures that bypass the repository ruleset entirely. For example, \texttt{Vouch}~\cite{vouch_ai_trust} reframes entry as a web-of-trust problem, where contributions are filtered based on endorsement before review: participants may need to be vouched for before interacting with selected project surfaces, and projects can extend this further into a web-of-trust model across repositories.
Similarly, \texttt{Agent-Trace}~\cite{agent_trace} proposes a machine-readable provenance format of AI contributions,
conversation links, and file- or line-level attribution, to shift the burden of ``proof of work'' from human disclosure to tool-level transparency.
These shifts suggest that the future of OSS governance lies not just in writing better markdown files,
but in reconfiguring the
infrastructure
itself to make GenAI contributions legible and governable by default.

\subsection{Implications for Maintainers and Platform Designers}

For \emph{maintainers}, the immediate implication is that there is no ``one-size-fits-all'' AI policy. Our taxonomy reads as a \emph{diagnosis-to-strategy map} (Figure~\ref{fig:strategy-orientation}): provenance or licensing risk points to Boundary Exclusion (A1) or Compliance and Provenance Safeguarding (A3); review overload points to Scope and Intentionality Control (C1), Verification and Evidence Gating (B2), or Capacity and Queue Control (C2); security-channel noise points to Security Channel Governance (C4); and agent- or platform-driven bypass points to AI Tooling Governance (B3) or Platform Reconfiguration (D1). Each path carries an explicit trade-off: stricter intake controls protect maintainer capacity but raise participation costs and may discourage external contribution, so projects should move up the map gradually rather than adopt maximal controls by default. The O2/O3 distinction matters for practice. Treating a review-throughput problem as an AI-specific one leads to over-engineering---new AI gates where existing quality controls suffice---whereas treating a provenance or accountability problem as generic quality control leads to under-engineering. Identifying the right orientation tells maintainers which family of strategies to use first, protecting the project without blocking valuable human--AI collaboration.

For \emph{platform designers}, the focus must shift from ``generation power'' to ``intake control'' and ``traceability.'' Shipping features that encourage ``effortless-to-produce but expensive-to-validate'' contributions without providing the tools to detect, prioritize, or block them creates an unsustainable burden on the community. Platforms must facilitate higher-information collaboration interfaces—such as machine-readable AI metadata—to ensure that the productivity gains of GenAI do not come at the cost of OSS community burnout.

\subsection{Future Work}

\emph{Efficacy of Disclosure Policies.} While \emph{Transparency and Disclosure} is a dominant and seemingly approachable strategy,
little is known about its actual compliance rate.
Future work may empirically measure the effectiveness of disclosure rules -- specifically, how often AI use is self-admitted versus hidden, and whether mandatory disclosure leads to systemic bias in how maintainers treat certain contributions.

\emph{Automated Triage.} As the volume of agent-assisted PRs increases, maintainers need proactive tools to manage the ``attention tax''~\cite{minh2026early}. Researchers may develop triage methods to identify low-quality agentic contributions with possibly high maintenance cost, enabling maintainers to fast-fail costly ``ghosting'' attempts.

\emph{Long-term Code and Community Health.} The rise of agentic coding may translate into a significant ``flow-debt tradeoff''.
He et al.~\cite{he2026speed} find that the adoption of Cursor can transiently boost development velocity, while leading to a persistent increase in static analysis warnings and code complexity.
Future longitudinal studies are needed to assess whether GenAI governance regimes successfully maintain long-term software quality, or if they inadvertently accelerate technical debt.

\section{Conclusion}

This paper provides a systematic account of how OSS projects govern GenAI in contribution workflows. Drawing on a corpus of public governance materials from 67 OSS projects, we showed how GenAI enters OSS collaboration at multiple workflow stages, what maintainer concerns it raises across these stages, and what governance responses projects articulate.

Our results show that GenAI governance in OSS is not reducible to the question of ``to ban or not to ban.'' Instead, projects confront governance challenges across the full contribution lifecycle -- spanning code reviewability, communication accountability, triage burden, security-channel integrity, provenance uncertainty, and platform constraints. We identified three governance orientations (prohibitionist, boundary-and-accountability, and quality-first) that explain how projects interpret these challenges differently, and abstracted 12 cross-project governance strategies and their policy instruments.

Taken together, these findings reveal a structural shift in OSS governance — from code review toward upstream admission control, and from text-based policies toward infrastructure-level assurance. For maintainers, the right governance path begins with diagnosing the bottleneck. Platform designers must invest in intake control and traceability to match the ease of generation.

\bibliographystyle{ACM-Reference-Format}
\bibliography{reference}

\end{document}